\documentclass[aps,prl,twocolumn,floatfix,floats,showpacs,superscriptaddress]{revtex4}
\usepackage{graphicx,latexsym}
\usepackage{dcolumn}
\usepackage{amsmath}

\begin{document}
\title{Nonadiabatic quantum pumping in mesoscopic nanostructures}
\author{C. S. Tang}
\affiliation{Physics Division, National Center for Theoretical
        Sciences, P.O.\ Box 2-131, Hsinchu 30013, Taiwan}
\author{C. S. Chu}
\affiliation{Department of Electrophysics, National Chiao Tung
University, Hsinchu 30010, Taiwan}
\begin{abstract}
We consider a nonadiabatic quantum pumping phenomena in a ballistic
narrow constriction.  The pumping is induced by a potential that has
both spatial and temporal periodicity characterized by $K$ and
$\Omega$. In the zero frequency ($\Omega=0$) limit, the transmission
through narrow constriction exhibits valley structures due to the
opening up of energy gaps in the pumping region --- a consequence of
the $K$ periodicity.  These valley structures remain robust in the
regime of finite $\Omega$, while their energies of occurrence are
shifted by about $\hbar\Omega/2$. The direction of these energy
shifts depend on the directions of both the phase-velocity of the
pumping potential and the transmitting electrons.  This frequency
dependent feature of the valley structures gives rise to both the
asymmetry in the transmission coefficients and the pumping current.
An experimental setup is suggested for a possible observation of our
nonadiabatic quantum pumping findings.
\end{abstract}
\pacs{73.23.-b, 73.23.Ad, 73.50.Mx, 72.10.-d}
\maketitle

The phenomenon of adiabatic quantum pumping in mesoscopic systems has
attracted much attention since the first proposal by
Thouless~\cite{Thouless83}.  The pumping phenomena refer to net
transport of charges at zero bias.  Mechanisms giving rise to quantum
pumping involves cyclic deformations of more than one parameter in the
system. The adiabatic deformation causes the transmission of a finite
amount of charges in one deformation cycle.  For the cases when this
finite charge transfer is quantized the pumping could be of importance
in establishing a standard of electric current~\cite{Niu90}. The first
experimental demonstration of the pumping of charges was reported by
Switkes {\it et al.}~\cite{Marcus99}. They applied sinusoidal voltages
to two metal gates that define the shape of an open quantum dot. The
phase difference between the two metal gates is an important adjustable
parameter for the pumping of charges.  Subsequent theoretical studies
on this pumping mechanism in quantum dots have invoked a double
oscillating barrier model~\cite{Wagner00,Wei00}.  More recently, it
was pointed out that quantum interference also plays an important role
in the pumping of charges~\cite{Zhou99}. Thus far, most of the studies
have concentrated on the adiabatic regime~\cite{Liu93}.  It is
legitimate then to explore the nonadiabatic aspect of the pumping
phenomena.

In this work, our purposes are threefold: to treat the pumping
phenomena nonadiabatically, to analyze in detail a novel pumping
mechanism found in our results, and to propose an experimental setup
for a possible realization of the mechanism. Towards these ends, we
have implemented a generalized scattering-matrix method that allows us
to solve the time-dependent Schr\"{o}dinger equation to effectively all
orders in the pumping potential~\cite{Tang00}. Hence our results are
not limited by the adiabatic approximation.  Furthermore, in analyzing
our results more closely, we are able to identify a pumping mechanism
that is associated with coherent inelastic scatterings of the
transmitting electron in the pumping region.  That the above quantum
transitions play a decisive role in this pumping phenomena shows the
nonadiabatic nature of the phenomena.  We stress that this pumping
phenomena do not require an asymmetry in the system configuration. The
phase velocity of the pumping potential alone is sufficient to cause
the pumping.

The pumping potential $V(x,t)$ we invoked acts upon a certain region of
a ballistic narrow channel. Apart from the overall spatial envelope of
the potential, $V(x,t)$ is periodic in time $t$, with a period
$T=2\pi/\Omega$, and periodic along the
longitudinal location $x$, with a period $L_{\rm p}=2\pi /K$.
We assume a simple envelope profile for the potential,%
\begin{equation}
 V(x,t)=V_0\cos (Kx-\Omega t)\Theta \left(L/2-|x|\right),
\end{equation}%
expecting it to have captured the essential physics of the pumping
mechanism.  Within the pumping region, the pumping potential has a
phase velocity $v = \Omega /K$, right-going in this case,  and the
longitudinal dimension it covered is $L$. As long as $L \gg L_{\rm p}$,
the detail form of the envelope profile --- such as a less abrupt
profile --- should not change the pumping features found in this work.
Meanwhile, similar potential form has been considered by O.
Entin-Wohlman {\it et al.\/}~\cite{Entin00} in their study on the
acoustoelectric  effect in a finite-length ballistic quantum channel.
The potential was generated from the surface acoustic wave, and was
assumed to be significant only inside the channel while it was totally
screened in the terminals. Even though the potential we consider in
this work is of similar form to that of theirs, the physical regimes of
interest are different. They considered the regime that incoherent
processes occur frequent enough in the channel to sustain a well defined
local distribution of the electrons~\cite{Entin00}.

Our interest, however, is in the coherent regime: that electrons
traversing the channel can maintain their phase coherence without
encountering any incoherent processes.  It turns out that the incoherent
and the coherent regimes exhibit different transport
characteristics.  For instance, in the incoherent regime, the pumping
feature was found to be most significant only in the vicinity of the
subband threshold~\cite{Entin00}, whereas in the coherent regime, we
find significant pumping features in the conductance plateau regions.

To simplify our presentation, we choose the energy unit $E^{*}=E_F$,
the length unit $a^*=1 / \! k_F$, the time unit $t^*=\hbar / E^*$, and
the frequency unit $\Omega^* = 1/t^*$. Here $E_F$ and $k_F$ represent,
respectively, the Fermi energy and the Fermi wave vector in the
reservoirs. The Hamiltonian is of the form ${\hat H} = {\hat H}_{y} +
{\hat H}_{x}(t)$, where ${\hat H}_{y} = -\partial^{2}/\partial y^{2} +
\omega_{y}^{2} y^{2}$ contains a transverse confinement, leading to
transverse subbands with energy levels $\varepsilon_{n} =
(2n+1)\omega_{y}$ and a subband level spacing $\Delta\varepsilon =
2\omega_y$. The time-dependent and the longitudinal part of the
Hamiltonian ${\hat H}_x (t)$ is of the form
${\hat H}_x (t) = -{\partial^2 /\partial x^2} + V(x,t)$.

At zero source-drain bias, the chemical potential $\mu$ is the same in
the reservoirs that connect to the two ends of the constriction. The
pumped current consists of contribution from
all electrons, given by%
\begin{equation}
I = - \frac{{2e}}{h}\int_0^\infty  {dE{\rm  }f(E - \mu )\left[ {T_ \to
(E) - T_ \leftarrow  (E)} \right]}\ . \label{I}
\end{equation}%
Here $f(E-\mu)$ is the Fermi distribution function  and $-e$ is the
charge of an electron. In addition, $T_ \to$ and $T_ \leftarrow$ are,
respectively, the total current transmission coefficients for the
right- and left-going incident electrons and for a given incident
energy $E$. Contributions from  all  sidebands, denoted by $m$, and
all subbands, denoted by $n$, have been taken into account, given by
\begin{equation}
T_{ \to ( \leftarrow )} (E) = \sum\limits_n {\sum\limits_m {^{'} T_{n
\to ( \leftarrow )}^m} } ,
\end{equation}%
where the primed summation includes only propagating components of the
transmitted electrons. The $T_{n \to ( \leftarrow )}^m$ denote the
transmission coefficients for the processes that an incident electron
in subband $n$, energy $E$, passes through the pumping region and ends
up in energy $E+m\Omega$ on the other end of the constriction.  These
coefficients are solved by a generalized scattering-matrix method that
we had established to treat time-dependent scattering potential
nonperturbatively~\cite{Tang96}. We had applied this method to cases of
longitudinally polarized field~\cite{Tang99} and transversely polarized
field~\cite{Chu96}.  Essentially, in this method, we segmented the
scattering potential along the longitudinal direction into slices,
solved the scattering matrix of each slice, and constructed the total
scattering
matrix~\cite{Tang00}. The energy derivative of the pumped current
$\partial I/\partial \mu$ is given, in the low temperature limit, by%
\begin{equation}
\frac{\partial I}{\partial \mu}  = - \frac{{2e}}{h} \left[ T_ \to(\mu)
- T_ \leftarrow (\mu)  \right].
\end{equation}%

In our numerical examples, we present the $\mu$ dependencies of the
current transmission coefficients and $\partial I/\partial \mu$.  The
values of the parameters used are consistent with that for a typical
high mobility two-dimensional electron gas formed at a ${{\rm
GaAs-Al}_{x}{\rm Ga}_{1-x}{\rm As}}$ heterostructure, with $E^{*} = 9\
{\rm meV}$, $a^{*} = 1/k_{{\rm F}} = 79.6$ ${\rm \AA}$, and
$\Omega^{*} = E^{*}/\hbar = 13.6\ {\rm Trad/s}$. We also choose
$\omega_y = 0.007$ such that the subband energy-level spacing
$\Delta\varepsilon = 0.014\ (\simeq 0.126\ {\rm meV})$. For the pumping
potential, we chose $V_0 = 0.002$, $K=0.15$, and the longitudinal range
$L=150$ such that $KL/2 \pi \cong 3.58$.  The
chemical potential $\mu$ is replaced by%
$$ X = {\mu\over \Delta\varepsilon} + {1\over 2},$$%
where the integral value of $X$ stands for the number of propagating
subbands in the constriction. The dependence of the current
transmission coefficients $T_\to$ and $T_\leftarrow$ on $X$ are
depicted by the solid and the dashed curves in Fig. 1, respectively.
The dotted curves are for $V_0 = 0$ and angular frequency $\Omega$
equals $0.2\Delta \varepsilon$, $0.6 \Delta \varepsilon$, and $\Delta
\varepsilon$, respectively, in Figs. 1(a)-(c).
\begin{figure}
\includegraphics[width=0.36\textwidth,angle=0]{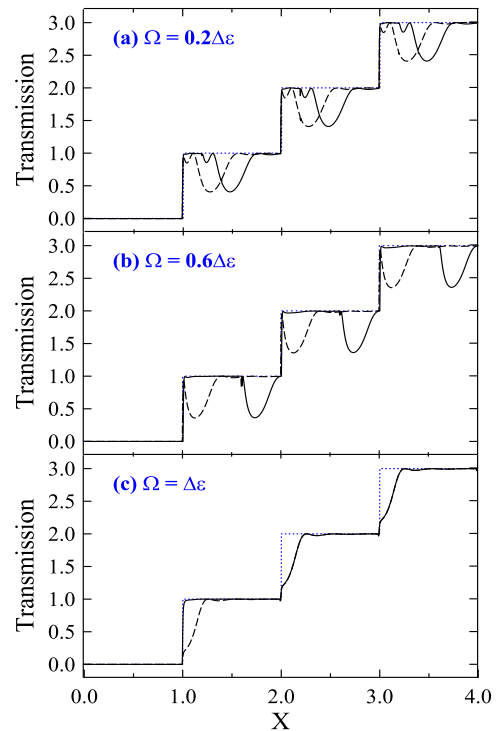}
\caption{ The current transmission coefficient as a function of $X$
for $\Omega/\Delta\varepsilon = 0.2, 0.6,$ and $1.0$ in Figs.\
1(a)-1(c), respectively. Other parameters are $K=0.15$, and $L=150$.
Solid curves and dashed curves are the transmissions of the right-
and left-going incident electrons, respectively, and for
$V_{0}=0.002$.  Dotted curves are for $V_{0}=0.0$.
}%
\label{fig:1}
\end{figure}%

The general features presented in Fig. 1 are summarized in the
following.  In the absence of the pumping potential, the transmission
coefficients $T_\to$ and $T_\leftarrow$ exhibit the well understood
quantized features, reflecting the number of propagating subbands in
the constrictions.  That both of the transmission coefficients are
given by the same dotted curve reflects the symmetry in the
transmission of the electrons with respect to their incident
directions.  This symmetry, however, is broken by the introduction
of pumping potentials to the constriction.  Upon the action of the
pumping potential, a valley structure is developed in each of the
plateau regions, but the energies of occurrence for these valley
structures are different for $T_\to$ and $T_\leftarrow$.   A valley
structure in Fig. 1 is characterized by a substantial drop in the
transmission coefficient, with maximum drop $\left| {\Delta T_{{\rm
max}} } \right| > 0.5$, and a large energy width, with $\Delta X_{\rm
Valley} \approx 0.4 \Delta \varepsilon$.  Valley structures for
$T_\leftarrow$ occur at lower energies than that for $T_\to$ on the
same plateau.   Interestingly, the separation in energy, or $\Delta X$,
between the minimums of the valley structures and belonging to the same
plateau for $T_\to$ and $T_\leftarrow$, increases with the angular
frequency $\Omega$ of the pumping potential.  In fact, our numerical
results show that $\Delta X = \Omega /\Delta \varepsilon$.   For the
case when $\Omega  = \Delta \varepsilon $, the separation becomes so
large that the $T_\to$ valley structures overlap with the
$T_\leftarrow$ valley structures of the next higher subbands.  These
valley structures and the asymmetry in the transmission coefficients
are the key results in this work.

We note in passing that there are
small oscillatory structures in Fig. 1(a) in the energy range between
the subband threshold and the valley structure.
These harmonic structures are associated with multiple scatterings
of the transmitting electrons in between the two edges of the pumping
potential.

Before we present a physical explanation for the valley structures, we
want to make one more remark: all the valley structures in Figs. 1(a)
and 1(b) are almost identical in their profile regardless of their
differences in the energies of occurrence and the incident direction of
the electrons. Furthermore, even though the valley structures in Fig.
1(c) are truncated at the threshold energies, the remaining profile
compares well with those in Figs. 1(a) and 1(b).  These
$\Omega$-independent features prompted us to check the $\Omega=0$ case,
and we find indeed the same valley structure profiles except that the
transmission symmetry is restored, with both $T_\to$ and $T_\leftarrow$
curves falling on top of one another. This $\Omega=0$ result is not
presented here, but we learn from this result that the energies of
occurrence for the finite $\Omega$ valley structures of $T_\to$ and
$T_\leftarrow$ have been shifted by about $\Omega/2$ and
$-\Omega/2$, respectively, from their $\Omega=0$ counterparts.  The
$\Omega=0$ valley structures are certainly associated with the
formation of an energy band gap inside the region acted upon by the
potential $V(x)$.  It is conceivable then that the valley
structures in the finite $\Omega$ regime are caused by the
establishment of a resonant coupling between some {\it degenerate}
states via the pumping potential.  To drive this point home and
to obtain an analytic expression for the energies of occurrence of the
valley structures, we propose a two-component approximation for the
wavefunction along the constriction, given by
\begin{eqnarray}
\psi (x,t) & = & \exp (ikx-iEt)  \nonumber \\
           &   & + C\exp [i(k - K)x-i(E-\Omega)t].
\end{eqnarray}
The above wavefunction has taken up a near resonance approximation
that describes the resonant coupling of an electron with photons.
Substituting Eq.\,(5) into a 1D Schr\"{o}dinger equation
$$\left[-\partial^{2}/\partial x^{2}+V_{0}\,{\rm cos}(Kx-\Omega\,t)
\right]\,\psi(x,t)=i\partial/\partial t\,\psi(x,t),$$
and retaining only terms of the form as in $\psi(x,t)$, we obtain
\begin{equation}
E = {1 \over 2}\left[ \Omega + \varepsilon _k  + \varepsilon _{k - K}
\pm \sqrt {\left( \Omega + \varepsilon _{k-K}  - \varepsilon _k
\right)^2  + V_0^2 } \right]\ ,
\end{equation}
where $E$ is the energy spectrum of an electron coupled with photons in
the system, and $\varepsilon_{k}=k^2$.

The energy spectrum has an energy gap at the resonant coupling
condition given by $\varepsilon_{k}= \varepsilon_{k-K}\pm\Omega$.  The
plus, and minus, sign in the resonant coupling condition is for
positive, and negative, $k$, respectively.  This in turn corresponds to
right- and left-going states. The 1D kinetic energy $\varepsilon_{k}$
at the center of the energy gap is $\varepsilon_{\rm Gap}=[(K/2)(1\pm
\Omega/K^2)]^2$, which can be approximated by $K^2/4\,\pm \Omega/2$
when $\Omega\ll K^2$.  In the case for a constriction, the energies of
occurrence will be at $X_{\rm Gap} =N+\varepsilon_{\rm
Gap}/\Delta\varepsilon$.  These $X_{\rm Gap}$'s correspond to the
minimums of the valley structures in Figs. 1(a)-(c). More specifically,
in Fig. 1(a), $X_{\rm Gap}=1.51, 2.51, 3.51$ for the right-going
electrons, and $X_{\rm Gap}=1.31, 2.31, 3.31$ for the left- going
electrons.  The matches of these values with the minimums of the valley
structures are remarkable.

Besides the energies of occurrence, Eq. (6) also provides us an
estimate for the energy gap, given by $\Delta X_{\rm
Gap}=V_{0}/\Delta\varepsilon=0.143$ for our case.  This $\Delta X_{\rm
Gap}$ is less than $\Delta X_{\rm Valley}\approx 0.4$, the width of the
valley structures in Fig. 1.  We have tried other cases of longer $L$
and find out that $\Delta X_{\rm Valley}$ decreases with increasing
$L$.  Its value becomes $0.2$ at $L=400$.  In addition, the maximum
drop in the transmission coefficient $\left| {\Delta T_{{\rm max}}}
\right|$ for the valley structures is very close to unity when
$L=400$.  Thus we conclude that the seeming discrepancy between
$\Delta X_{\rm Gap}$ and $\Delta X_{\rm Valley}$ in Fig. 1 is resulted
from the finite $L$ effect.  The detail profile of the transmission
coefficients, however, is given by our nonperturbative approach.

We have shown that the valley structures in the finite $\Omega$
regime are resulted from the coherent inelastic scatterings in the
pumping region.  The $\Omega$-dependence in the energies of occurrence
$X_{\rm Gap}$ reflects the breaking of the transmission
symmetry by the phase velocity of the pumping potential.

\begin{figure}
\includegraphics[width=0.36\textwidth,angle=0]{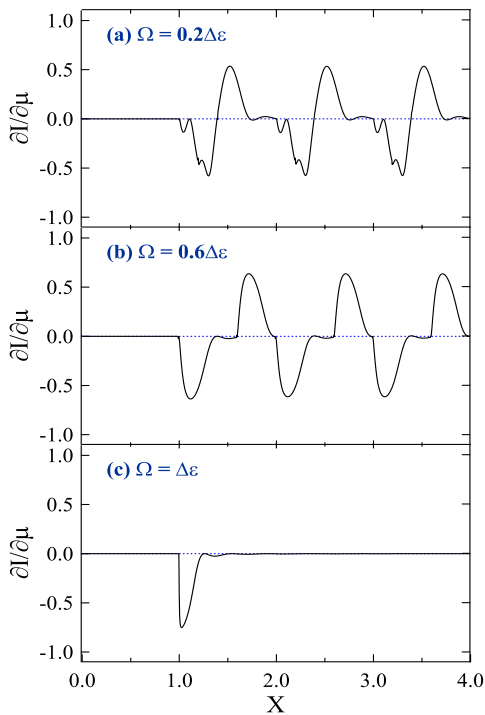}
\caption{ The energy derivative of the pumped currents, in units of
$2e/h$, as a function of $X$ for $\Omega/\Delta\varepsilon = 0.2,
0.6,$ and $1.0$ in Figs.\ 2(a)-2(c), respectively. Other parameters
are the same as those in Fig.\ 1.  Solid curve is for $V_{0}=0.002$
and dotted curves are for $V_{0}=0.0$.
}%
\label{fig:2}
\end{figure}%

In Fig.\ 2, we present the effects of the transmission asymmetry
on the quantum pumping by plotting $\partial I/\partial \mu$ as a
function of $X$. The parameters
are the same as those in Fig.\ 1.  Major features shown in Figs. 2(a),
2(b) are alternate valley and broad peak structures.  These are due
to suppression in the transmission for electrons incident from the
right, and left, electrode, respectively.  The valley and the broad
peak structures barely resemble each other in Fig. 2(a).   However, the
corresponding structures in Fig. 2(b) are almost inversion images of
one another.  Since the valley structures for the two
transmission coefficients overlap in Fig. 1(a), their contributions
to $\partial I/\partial \mu$ suffer from partial cancellation between
themselves.  This cancellation hampers the resemblance between the
valley and the broad peak sturctures in Fig. 2(a).
On the other hand, the valley structures are well separated in
Fig. 1(b), thus preserving the resemblance between the
valley and the broad peak structures in Fig. 2(b).  The case of exact
cancellations is presented in Fig. 2(c), when the valley structures of
$T_{\to}$ overlap with the valley structures of $T_{\leftarrow}$ of the
next subbands.  Only the first valley structure survives and the
net flow of the electrons is in the same direction as the phase velocity
of the pumping potential.  Finally, we stress that the pumping effect is
most significant in the conductance plateau regions.  This is in sharp
contrast to the pumping effect in the incoherent regime~\cite{Entin00}.

For the possible realization of the pumping mechanism proposed in
this work, we suggest an experimental setup that has taken full
advantage of the recently developed finger gate technology
~\cite{Liang98,Tralle01}.  The basic structure is a narrow constriction
defined out of a two dimensional electron gas by a pair of split-gates
(SG).  On top of this SG, and separated vertically by an insulating
layer of submicron thickness, are two interdigitated finger gates (FG)
that are oriented transversely.  Each of these FGs has a period
$L_{\rm{p}}$.  The pumping potential could be
generated by subjecting the FGs to AC bias of the same
frequency while maintaining a phase difference between the two FGs.
The control of the phase difference in the bias
voltage between two neighboring gates was successfully demonstrated
by Switkes {\it et al}~\cite{Marcus99} in the $f\approx 10$ MHz region.
More recently, phase-shifter operating in the $f\approx 10$ GHz region
has been fabricated using a 0.6-$\mu$m GaAs process~\cite{Fran01}.
Hence
we believe that the suggested experimental setup, though poses a
stringent challenge to the experimentalists, is within reach of the
present nanotechnology.

In conclusion, we have proposed and have analyzed in detail a
nonadiabatic quantum pumping mechanism.  We have demonstrated
the robustness of such pumping mechanism due to its resonant
coupling nature.  And we have proposed an experimental setup for
the possible realization of such mechanism.

C.S.T.  thanks Professor Y. C. Lee for his enthusiastic
encouragement. This work was supported by NSC of the ROC under
Contract No. NSC89-2112-M-236-001 and NSC90-2112-M-009-003.

\end{document}